\documentclass[journal]{IEEEtran}
\pdfoutput=1
\usepackage{graphicx}

\begin{document}
\title{Fabrication of Magneto-Optical Atom Traps on a Chip}

\author{G. Lewis, Z. Moktadir, C. Gollasch, M. Kraft, S. Pollock, F. Ramirez-Martinez, J. P. Ashmore, A. Laliotis, M. Trupke, E. A. Hinds,\,~\IEEEmembership{Fellow,~OSA,} \thanks{This work was in supported by the EU SCALA project, the UK EPSRC and the Royal Society.} \thanks{We are indebted to Mr J. Dyne for expert technical assistance.} \thanks{G. Lewis, Z. Moktadir, C. Gollasch and M. Kraft are with the
School of Electronics and Computer Science, University of Southampton,
Southampton,SO17 1BJ, United Kingdom (email:Gnl04r@ecs.soton.ac.uk)} \thanks{S. Pollock, F. Ramirez-Martinez, J. P. Ashmore, A. Laliotis,
M. Trupke \& E. A. Hinds are with the Centre for Cold Matter,
Imperial College, Prince Consort Road, London, SW7 2AZ, United
Kingdom (email:ed.hinds@imperial.ac.uk)}}

\maketitle

\begin{abstract} Ultra-cold atoms can be manipulated using microfabricated
devices known as atom chips. These have significant potential for
applications in sensing, metrology and quantum information
processing. To date, the chips are loaded by transfer of atoms from
an external, macroscopic magneto-optical trap (MOT) into microscopic
traps on the chip. This transfer involves a series of steps, which
complicate the experimental procedure and lead to atom losses. In
this paper we present a design for integrating a MOT into a silicon
wafer by combining a concave pyramidal mirror with a square wire
loop. We describe how an array of such traps has been fabricated and
we present magnetic, thermal and optical properties of the chip.
\end{abstract}


\section{Introduction}
Atom chips are microfabricated devices that control
electric, magnetic and optical fields in order to trap and
manipulate cold atom clouds \cite{Hinds:1999}, \cite{Folman:2002},
\cite{Fort:2007}, \cite{Eriksson:1} and to form Bose-Einstein
condensates \cite{Hansel:2001}, \cite{Ott:2001},
\cite{Sinclair:2005}. Potential applications include atomic clocks
\cite{Knappe:2005}, atom interferometers \cite{Shin:2005},
\cite{Schumm:2005}, and quantum information processors
\cite{Trupke:2007}, \cite{Schmiedmayer:2002}. Silicon is one of
several materials used as a substrate for atom chips. It is
attractive for this purpose because its properties are well-known
and fabrication techniques are highly developed. The small scale of
microfabricated current-carrying wires makes it easy to generate
strong magnetic field gradients near the surface of the chip,
forming tight traps for paramagnetic atoms. The loading of such
magneto-static traps usually starts with a magneto-optical trap
(MOT) typically some 3-4\,mm from the surface. This collects atoms
from a tenuous, room-temperature vapour and cools them, typically to
100\,$\mu $K, using circularly polarised light beams in conjunction
with a spherical quadrupole magnetic field. The atoms are sometimes
further cooled to a few tens of $\mu $K using optical molasses,
before being captured in a weak magnetic trap to form a large atom
cloud, typically 1\,mm in size. At this point, the atoms still have
to be handed over to the microscopic magnetic traps on the chip, a
process that involves further compression of the cloud and very
accurate positioning of the atoms. This sequence of loading and
transfer is complicated and could be largely eliminated if the MOT
were integrated into the chip. Moreover, integration would open up
the possibility of building arrays of MOTs to prepare large numbers
of independent cold atom clouds.

This paper describes the fabrication and initial testing of an
integrated array of MOTs on an atom chip, as proposed by Trupke
\textit{et al.}\,\cite{Trupke:2006}. Each of these MOTs automatically
prepares all the required light beams from a single circularly
polarised input beam by reflecting the light in a concave square
pyramid of mirrors \cite{Lee:1996}. This greatly reduces both the
number of expensive optical components needed to prepare the light
beams and the amount of laser power needed. Integrated wires
encircling the opening at the base of the pyramid produce the
required magnetic field distributions with modest electrical power
consumption and accurate positioning. The fabrication of an
integrated MOT array on a chip represents an important step towards
a truly integrated atom chip for portable applications.

The atom chip we have fabricated has 6 rows of pyramids, ranging in
size from 200\,$\mu $m to 1200\,$\mu $m, serviced by 12 separate
wires to produce the magnetic fields. For pyramids up to 600\,$\mu
$m, the encircling wires have a width of 25\,$\mu $m. The larger
pyramids are serviced by wires of 50\,$\mu $m width. In total there
are 48 pyramid MOTs. The whole chip is packaged into a ceramic pin
grid array (CPGA) with multiple wire bonds to bring high currents in
and out of the chip. The silicon sidewalls of the pyramids are
coated with gold to create micro-mirrors for reflecting the laser
light. The pyramids formed by etching silicon have a 70.5$^{o}$ apex
angle, rather than the ideal 90$^{o}$. The optical properties of
such a pyramid have already been investigated in \cite{Trupke:2006}.
In those experiments it was observed that light reflected near the
diagonal edges can prevent the MOT from working. Here we have
developed the necessary fabrication steps to eliminate these
reflections by removing the gold near the corners of the pyramids.

This paper is organised as follows. Section\,\ref{Sec:principles}
outlines the principles of atom trapping in these pyramidal
micro-mirrors, section\,\ref{Sec:fabrication} describes the
microfabrication, section\,\ref{Sec:characterisation} presents
initial tests of the device and section\,\ref{Sec:outlook} discusses
some prospects for using the chip in applications.

\section{Principle of the MOT on a chip}\label{Sec:principles}
Circularly polarised light, incident along the axis of a square
pyramid, is reflected by the four metal mirrors that form the
pyramid. At each reflection the helicity of the light is reversed.
If the pyramid has a 90$^{o}$ angle between opposite faces, these
reflections produce three counter-propagating pairs of light beams
that are mutually orthogonal. Together with a magnetic quadrupole
field, this configuration creates a MOT, whose radiation pressure
forces cool and trap atoms from a room-temperature vapour
\cite{Lee:1996}, \cite{Lindquist:1992}.

\begin{figure}
\centering
\includegraphics{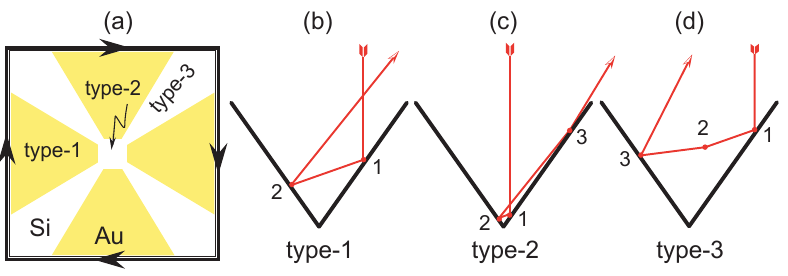}
\caption{(a) Plan view of pyramid showing the three regions of
reflection and the encircling current. The following three figures
illustrate the sequence of reflections. (b) Cross section through
pyramid showing a type-1 reflection. (c) Type-2 reflection. (d)
Type-3 reflection, where the ray is intercepted by the adjacent
mirror at the point marked 2.} \label{Fig:rays}
\end{figure}

We form the pyramids by etching a silicon wafer, cut on the
{\{}100{\}} plane. Potassium hydroxide (KOH) etches anisotropically
through square openings to reveal the {\{}111{\}} planes, which form
hollow pyramids with apex angle $70.5^{o}$. This departure from a
right angle causes the beams to be reflected into a variety of
directions. We classify these beams as Type 1, 2, or 3, according to
the region of the pyramid where the first reflection occurs, as
shown in Fig.\,\ref{Fig:rays}a. Type-1 rays are reflected on two
opposite sides of the pyramid before leaving, as illustrated in
Fig.\,\ref{Fig:rays}b. Type-2 rays also reflect on opposite faces,
but strike the original face again before leaving, as shown in
Fig.\,\ref{Fig:rays}c, making a total of three reflections. After
the first reflection in Fig.\,\ref{Fig:rays}d, type-3 rays head
towards the opposite face, but because they are incident close to
the diagonal edge of the pyramid, they are intercepted on the way by
the adjacent face. Here they undergo a grazing reflection, marked
(2) in the figure, where the helicity of the light is reversed.
Finally, the opposite face is reached for a third reflection.

\begin{figure}
\centering
\includegraphics{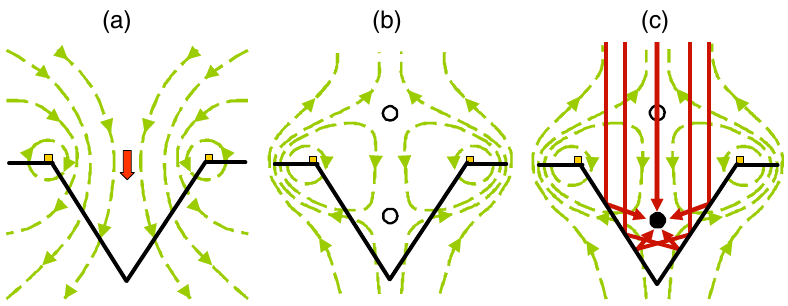}
\caption{(a) Dashed lines: magnetic field created solely by the
wires around pyramid opening. Block arrow: maximum field strength is
at the centre of the loop. (b) With the addition of a uniform bias,
the net field acquires two minima, indicated by solid circles.  (c)
Circularly polarised light creates a magneto-optical trap at the
field minimum indicated by a filled circle.  The beams that
contribute are shown as solid arrows.} \label{Fig:MOT}
\end{figure}

The wires electroplated on our chip form a square loop of side $L$
around the pyramid base. A current $I$ in this loop makes a magnetic
field as shown in Fig.\,\ref{Fig:MOT}a. At the centre of the loop,
in the plane of the wire, the field points downward with a magnitude of
$\;\textstyle{{2\sqrt 2 } \over \pi }\mu _0 I/L$. We
superpose a uniform vertical bias field in order to create the
quadrupole field configuration required by the MOT, as shown in
Fig.\,\ref{Fig:MOT}b. The strength $B_{T}$ of this bias is chosen to
centre the quadrupole halfway between the base and the apex, at a
distance $\textstyle{1 \over {2\sqrt 2 }}L$ from the surface of the
chip. This requires $B_T =\;\textstyle{{8\sqrt 2 } \over {3\pi \sqrt
5 }}\mu _0 I/L=0.54\mu _0 I/L$. The light beams needed to create the
magneto-optical trap inside the pyramid are formed from the incoming
circularly polarised beam by the first and second reflections of the
type-1 rays, as shown in Fig.\,2c. The type-2 rays produce a slight
imbalance in the MOT force, but are largely unimportant. The type-3
rays tend to destabilise the MOT because they have the wrong
helicity and produce a strong force that pushes the atoms out of the
pyramid instead of trapping them.

We have conducted preliminary experiments using a large, glass
pyramid coated with gold. Initially, this failed to produce a MOT
because of the presence of the type-3 rays. Atoms were successfully
trapped once the gold was removed from the areas where the type-2
and type-3 rays are produced. We were also able to make the MOT
operate \cite{Trupke:2006} by using a lower-reflectivity (78{\%})
coating of aluminium, which was not cut away at the edges and
centre. In such a pyramid, the intensity of the harmful type-3 light
is decreased relative to the type-1 beams, because of the additional
reflection by the lossy surfaces.

The current in the wire determines the vertical gradient in the MOT,
which is related to the corresponding bias field according to
$\textstyle{{26\sqrt 2 } \over {15}}B_T /L=2.45B_T /L$. Under
typical operating conditions used in most MOT experiments, the
gradient is approximately 0.15\,T/m. This is readily achieved on the
chip because of the small scale: for example in the 1\,mm loop it
requires 0.1\,A. The same wires can also be used to create a purely
magnetic trap, with a depth of $\mu\,B_T $, where $\mu$ is the
magnetic moment of the atom. Such a trap can hold atoms for many
seconds provided its depth exceeds the temperature of the atoms by a
factor of 5 or so. Ideally, we would like to be able to hold a
$100\,\mu$K cloud, which requires a 0.7\,mT bias field,
corresponding to a current in a 1\,mm loop of 1\,A. The feasibility
of using such currents is discussed more fully in
Sec.\,\ref{Subsec:HeatingTests}. Further details about magnetic
trapping of neutral atoms can be found in Ref.\,\cite{Hinds:1999}.

\begin{figure*}[t]
\centering
\includegraphics{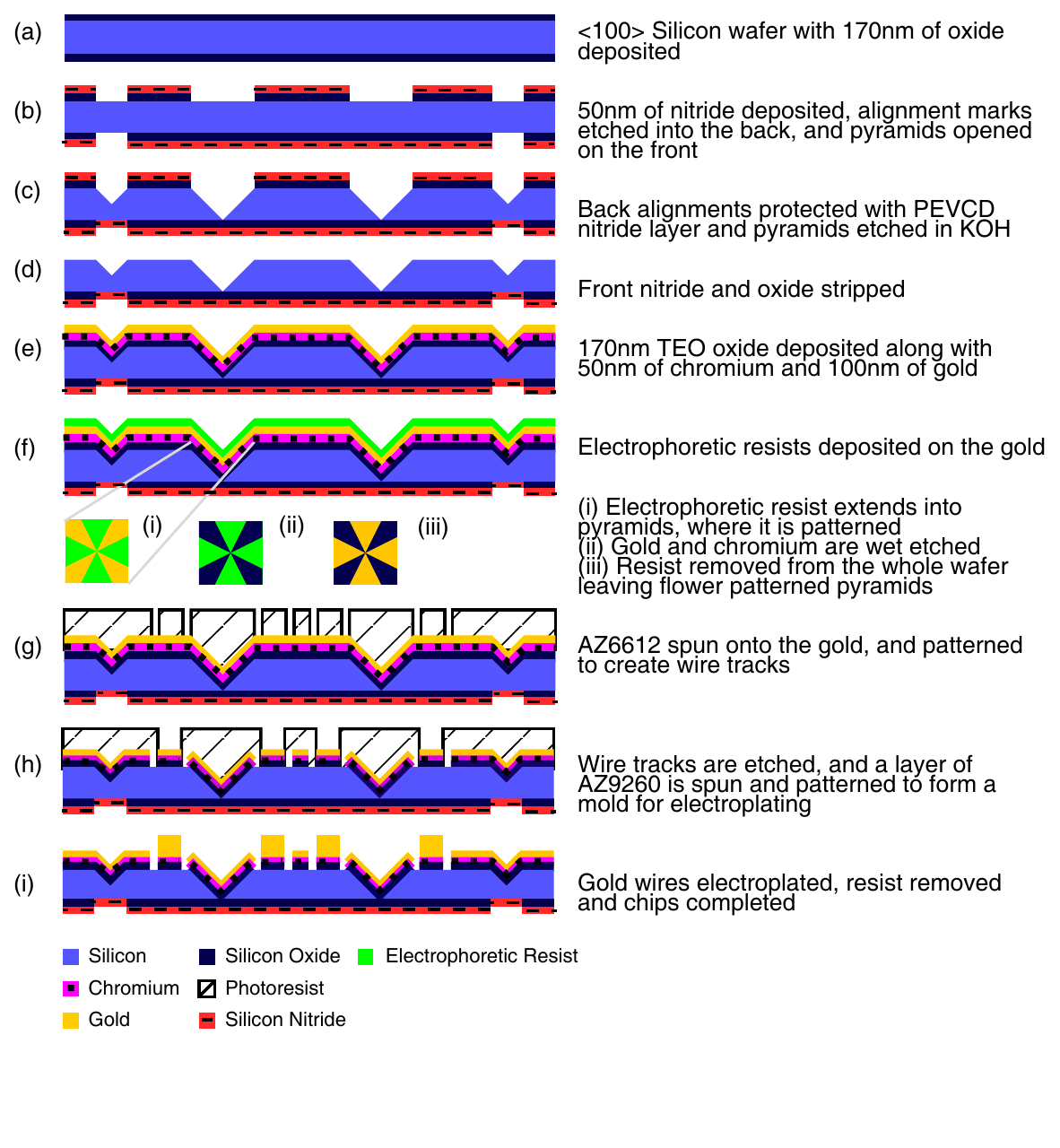}
\centering
\caption{Process flow for fabricating the pyramid MOT
chip.}
\label{Fig:flow}
\end{figure*}

In summary, the aims of the fabrication were to create pyramidal
micro-mirrors which would not reflect the damaging type-3 beams, and
to surround these mirrors with wires able to create the required
magnetic field strengths and gradients. In the following sections we
describe how these aims have been achieved.

\section{Fabrication}\label{Sec:fabrication}

The fabrication processing sequence is shown in
Fig.\,\ref{Fig:flow}. We start by preparing the wafer and etching
the pyramids. The whole surface is then coated with gold and
patterned to remove the gold from the type-3 regions and to form the
wire tracks. In these steps, the main fabrication challenge was to
cover the wafer with a uniform resist layer, because the pyramid
openings in the surface impeded the flow of the resist during
spinning. Finally, the wires are electroplated to make them thick
enough to carry the required current. These processes are described
in more detail in the next three sub-sections.

\subsection{Wafer preparation and pyramid etch}

The process begins with a 4 inch, 1mm-thick silicon wafer cut on the
\{100\} plane. This is given a standard RCA and fuming nitric
acid clean. A 170\,nm-thick layer of silicon dioxide is grown by wet
oxidation in a furnace at 1000\,C (Fig.\,\ref{Fig:flow}a).
Subsequently a 50\,nm layer of low stress silicon nitride is
deposited on both sides by LPCVD. A reactive ion plasma etch is then
used to make alignment marks on the back side of the wafer for
accurate positioning of all the masks used in the fabrication
process.

In order to make the pyramids, a 1\,$\mu$m layer of photo resist
AZ6612 is spun onto the wafers. An array of square openings is
patterned into the resist using a photo mask on a Karl Suss MA8
contact mask aligner. The silicon nitride and dioxide layers are
removed through these openings using a dry plasma etch and the
resist is then stripped in a plasma asher (Fig.\,\ref{Fig:flow}b).
Before etching the silicon, the backside alignment marks are
protected by depositing a layer of PECVD silicon nitride 1$\mu $m
thick and the wafer edges are protected by PTFE tape. The wafer is
then etched for 19 hours in KOH at a concentration of 33{\%} by
volume and at a temperature of 80\,C. This produces pyramidal pits
bounded by the four most slowly etched surfaces $\{1,1,1\}$,
$\{\overline{1},1,1\}$, $\{1,\overline{1},1\}$ and
$\{\overline{1},\overline{1},1\}$ (Fig.\,\ref{Fig:flow}c). The faces
of the pyramids are very smooth because of the layer-by-layer
etching mechanisms involved \cite{Moktadir:1997}
and have been shown to have rms surface roughness as low as 0.5 nm
\cite{Trupke:2006}. The wafers are once again cleaned in a fuming
nitric acid bath before the remaining silicon nitride is stripped
from the front by a dry plasma etch and the remaining silicon
dioxide is removed in an HF dip (Fig.\,\ref{Fig:flow}d). Finally,
the whole front surface is covered with a plasma-enhanced TEOS oxide
layer, 170 nm thick. This is to provide electrical insulation
between the silicon and the metallic coating that comes next.

\subsection{Metallic coating and flower patterning }
The metal coating consists of 50\,nm of chromium and 100\,nm of
gold, evaporated onto the front of the wafer
(Fig.\,\ref{Fig:flow}e). In order to make the flower pattern on the
faces of the pyramids, we require them to be covered by a uniform
layer of resist. Spinning the resist at this stage does not produce
a uniform layer because of the large depth of the pyramidal pits. In
order to avoid this problem we use electrophoretic deposition of
Eagle 2100 negative photo resist. The wafer is placed in the resist
bath and heated to 33\,C, where it acts as a cathode at -125\,V. It
remains in the bath until the current drops to zero, then it is
rinsed in de-ionised water and dried in a vacuum oven at 65\,C. At
this point, the resist remains tacky, so the wafer is dipped into
Eagle 2002 topcoat for 30 seconds and again dried in the vacuum
oven. This method leaves a highly uniform layer of resist over the
complex topography (Fig.\,\ref{Fig:flow}f).
\begin{figure}
\centering
\includegraphics{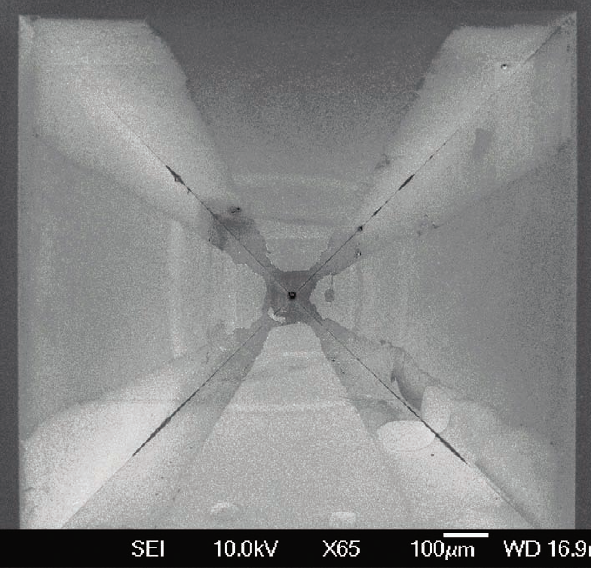}
\caption{SEM image showing the flower pattern created on the side
walls of all the pyramids.}\label{Fig:SEM}
\end{figure}

Next, the wafer is exposed to ultraviolet light for 77\,s at
6.5\,mW/cm$^{2}$, using a mask designed to remove the gold from the
type-3 regions. This pattern is developed using Eagle 2005 developer
heated to 40\,C, which is sprayed onto the surface for 4 minutes.
The wafer is then dipped for 1 minute in water at 80\,C  to remove
the resist residues and to smooth the surface of the resist. It is
dried in a vacuum oven and de-scummed for 3 minutes in oxygen plasma
at 110\,C (Fig.\,\ref{Fig:flow}f(i)).

The exposed gold and chromium are removed by a 35\,s potassium
iodide etch, followed a 5\,s chrome etch
(Fig.\,\ref{Fig:flow}f(ii)). The wafers are then sprayed for 15\,min
to strip the resist with Eagle 2007 remover at 50\,C, and placed in
an asher for 1\,hr at 600\,W and 110\,C
(Fig.\,\ref{Fig:flow}f(iii)). This process leaves a flower pattern
on the gold inside each pyramid, as shown in Fig.\,\ref{Fig:SEM}.
This scanning electron microscope image shows that the gold has been
removed in the diagonal and central regions corresponding to the type-2
and type-3 areas illustrated in Fig.\,\ref{Fig:rays}a. We found that it was difficult to achieve a uniform exposure of the resist inside the pyramid. When the lower part of the pyramid (near the apex) is correctly exposed, the upper part tends to be over-exposed. The dark lines on the diagonals and the dark spot at the apex are due to residual resist.  The colour of the silicon surface appears darker at increasing depth in the pyramid. We attribute this to charging of the silicon dioxide surface by electron bombardment from the SEM.

\subsection{Wire fabrication}
The final stage of fabrication is to form the wires around the base
of each pyramid, as described in \cite{Koukharenko:2004}. First, the
tracks are patterned by optical lithography on the gold coating,
then the wires are electroplated to provide the 3\,$\mu$m thickness
required for carrying the electrical current.

\begin{figure}
\centering
\includegraphics{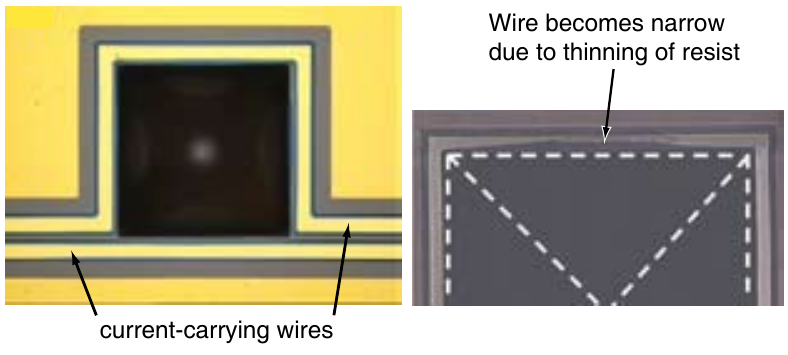}
\caption{Optical microscope image of the wire tracks. (a) The wires
are well formed around a $200\,\mu$m pyramid. (b) On the side
furthest from the axis of spinning, the wire of a 1.2\,mm pyramid is
malformed because the resist was too thin.}\label{Fig:tracks}
\end{figure}

Once again, spinning a uniform layer of resist is very challenging
since the radial flow is interrupted by the pyramid openings. To
solve this problem we use the viscous photoresist AZ4533 as follows.
A primer is spun onto the wafer to help adhesion. Then the wafer
needs to be completely flooded with the resist before the spinning
begins to ensure coverage inside pyramids and in the areas of the
spin shadow just outside the pyramids. Spinning is carried out at
500\,rpm for 10\,s, followed by 30\,s at 2000\,rpm. We create a
5\,$\mu$m layer of resist, which varies in thickness by
approximately $\pm $3\,$\mu$m where the pyramids interfere with the
flow (Fig.\,\ref{Fig:flow}g). The uniformity of the resist improves
substantially around pyramids that are smaller than 1\,mm, and
impeccable wires are formed around the smallest pyramids, as shown
in Fig.\,\ref{Fig:tracks}a. However, the thinning of the
photo-resist in the spin shadow of the larger pyramids leads to
overexposure and results in a thinning of the wire tracks there, as
shown in Fig.\,\ref{Fig:tracks}b.

A 6\,$\mu$m-thick electroplating mould is then created by optical
lithography of AZ9260 which is spun at 500\,rpm for 8\,s, followed
by 30\,s at 4000\,rpm. This layer is exposed at 200\,mW/cm$^{2}$ for
30\,s to create the plating mould with a minimum thickness of
3\,$\mu$m (Fig.\,\ref{Fig:flow}h). The electrochemical deposition is
controlled by an Autolab PGSTAT30. We use a commercial cyanide-free
gold plating solution (Metalor Technologies UK, with 1\,g of gold
per litre of solution). A standard three-electrode setup is used
with a platinum counter electrode. The deposition bath is placed in
a water bath kept at 50\,C and the solution is agitated throughout
the deposition process using a magnetic stirrer. The current is
fixed at 6\,mA, corresponding to a current density of
5\,mA/cm$^{2}$, and the plating is run for 15 minutes to achieve a
3\,$\mu$m-thick gold deposit (Fig.\,\ref{Fig:flow}i). Finally, the
resist is stripped to leave the free-standing gold wires.

Electrophoresis could potentially provide a solution to the problem
of the thinning of the wires. However, it can only be used when
pattering the wire tracks and not when creating the electroplating
mould, since the Eagle 2100 resist can only cover areas previously
covered with gold. For this reason the fabrication of the wires was
entirely designed and performed with the use of spin-on positive
photoresists.

\subsection{Packaging}

\begin{figure}
\centering
\includegraphics{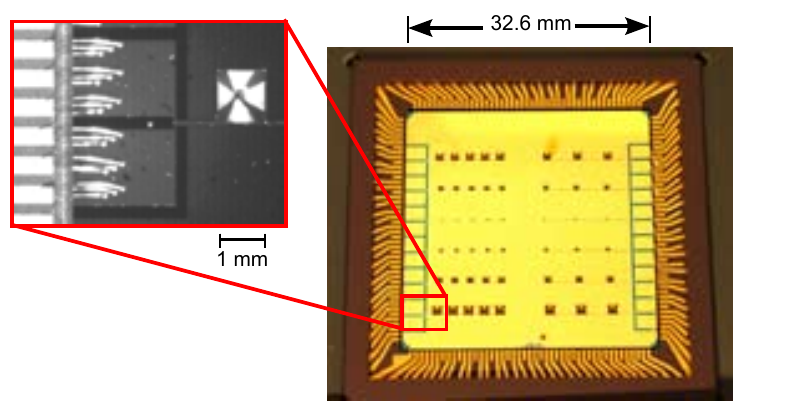}
\caption{Photographs of the pyramid-MOT chip mounted in its ceramic
pin grid array package. Right: the the chip in its package.  Left:
detailed view of bond wires connecting the package to the chip wire
pads, and a pyramid with the flower-patterned gold
coating.}\label{Fig:mounted}
\end{figure}

Figure\,\ref{Fig:mounted} shows on the right an array of patterned
pyramidal micro-mirrors with integrated current-carrying wires. The
chip is fixed to a ceramic pin grid array (CPGA) package (CPG18023,
Spectrum Semiconductor Materials Inc.) using an epoxy compatible
with ultra-high vacuum (Bylapox 7285). Gold bond wires connect the
package pins to the microfabricated chip wires, with 9 bond wires
serving each end of a chip wire. The view on the left of
Fig.\,\ref{Fig:mounted} shows the bond wires connecting the chip to
the CPGA. The ceramic package is then plugged into a pin grid array
(PGA) of sockets, which in turn is soldered into an FR4 printed
circuit board. This design permits a large number of connections to
be made in the limited space available in the high-vacuum chamber.

We place the chip assembly in a stainless steel vacuum chamber where
a 20\,l/s ion pump yields a pressure below the 10$^{-9}$mbar limit
of our gauge after modest baking at 100\,C for 2 days. At this low
pressure, a trapped atom will remain undisturbed by the background
gas for typically 10\,s.

A stray magnetic field moves the zeros of magnetic field that form
the centre of each MOT. To ensure that every pyramid encloses a
field zero, we require the stray field to be less than 10\,$\mu$T
over the chip. Much stronger fields were initially found in the
recess of the CPGA where the wafer is mounted, due to permanent
magnetisation of the nickel/gold coating. We therefore removed the
coating by sandblasting. Immediately above the Kovar pins, there
remained a stray field that reached a maximum of 10\,$\mu$T, but
this was of short range and did not affect the 3\,cm~x~3\,cm active
area of the chip. There the field was below 500\,nT, the gradients
were similarly negligible, and remagnetising the assembly produced
no perceptible change at that level.

\section{Device characterisation}\label{Sec:characterisation}
In this section we characterise the magnetic properties of our
device, demonstrating that the micro-fabricated wires can carry
sufficient current for magneto-optical and purely magnetic trapping.
Measurements of the heating of our chip wires allow us to infer the
maximum depth and gradient achievable in our trap. Finally, it is
shown that the flower patterning of the pyramid coating leads to the
desired suppression of type-3 beams.

\subsection{Magnetic field measurements}

\begin{figure}
\centering
\includegraphics{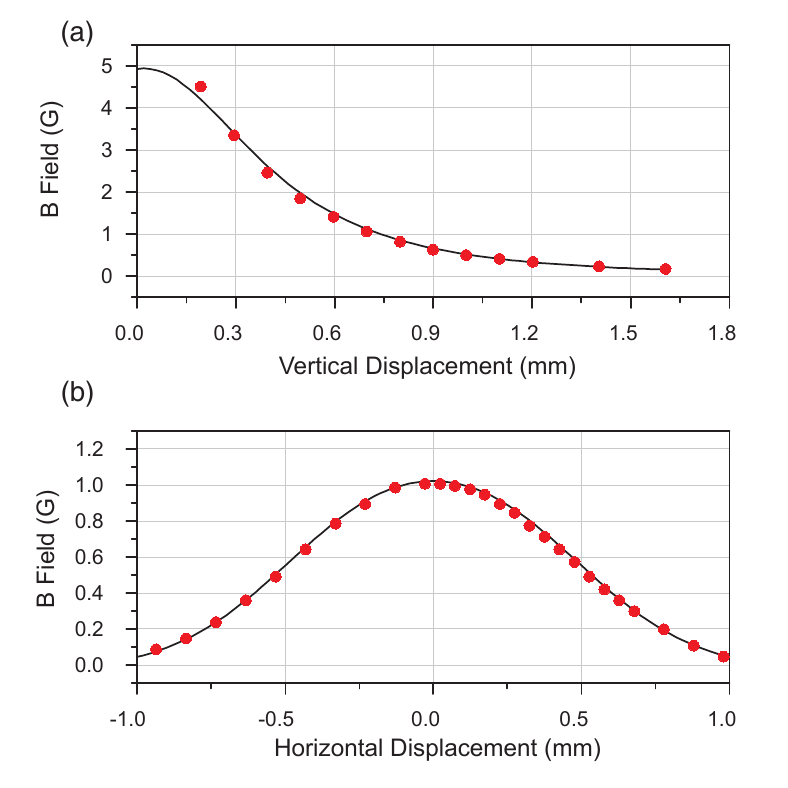}
\caption{Vertical component of magnetic field above an 800\,$\mu$m
pyramid with 480\,mA in the wire. Dots: Hall probe measurement.
Lines: Theory. (a) Vertical variation along the pyramid axis. (b)
Transverse variation parallel to the pyramid row, through the
pyramid axis at a height of 0.7 mm.} \label{Fig:HallScan}
\end{figure}

The wires around the base of each pyramid almost form a square loop,
as shown in Fig.\,\ref{Fig:tracks}a. In order to calculate the
magnetic field of the MOT, it is adequate to approximate this by a
fully closed square loop. The solid lines in Figs.\,7a and b show
the vertical field component calculated for a current of 480\,mA in
wire 50\,$\mu$m wide, centred on a 895\,$\mu$m square. This
corresponds to the geometry of the 800\,$\mu$m pyramid. We measured
the field above this pyramid using a Hall probe (Lakeshore 421
Gaussmeter), as shown by the circles in Fig.\,\ref{Fig:HallScan}. On
the left is a vertical scan along the axis of the pyramid, while the
scan on the right is horizontal at a height of 0.7\,mm. The two are
entirely consistent.

\subsection{Heating tests}\label{Subsec:HeatingTests}
Although the gold wires are very good conductors, they do of course
dissipate electrical energy and, at sufficiently high currents, they
blow like a fuse. Considerably below that limit, the wafer heats up
to 120\,C. At this temperature the epoxy used to bond the silicon to
the CPGA begins to decompose, losing strength and outgassing
strongly, thereby compromising both the mechanical stability of the
chip and the high vacuum. This limiting temperature determines the
largest current that we run through the wires, and thereby limits
the field and field gradient that are available.

\begin{figure}
\centering
\includegraphics{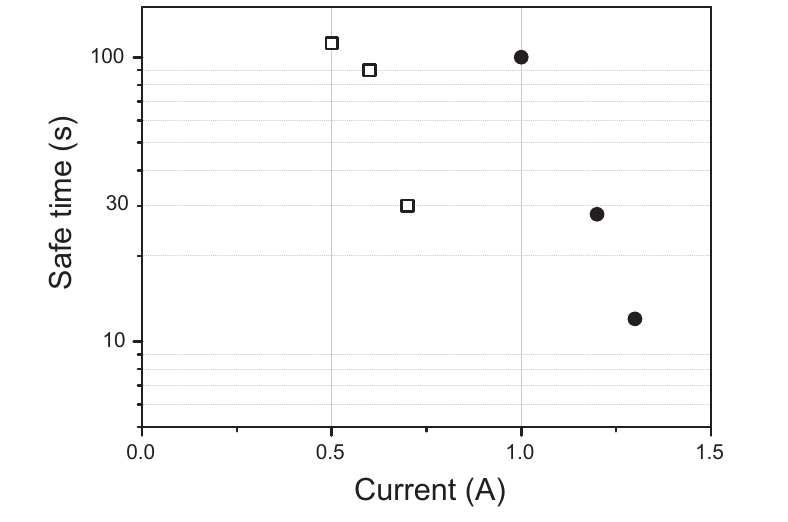}
\caption{Time taken for the chip wire temperature to reach 120\,C
for various currents.  Filled circles: 50\,$\mu$m-wide wires. Open
squares: 25\,$\mu$m-wide wires.}\label{Fig:heating}
\end{figure}

With the chip mounted and placed under moderate vacuum (10$^{-5}$
mbar), we first tested how much current could be passed through
individual bond wires 50\,$\mu$m in diameter and 2-3\,mm in length.
We found that these blow at approximately 1.8\,A, but can survive
indefinitely at 1.5\,A. Since each chip pad is normally connected by
9 wires in parallel, failure of the bond wires is not a limiting
factor. In order to determine the operating currents for the chip
wires, we monitored the temperature of the assembly at several
points using thermocouples and we monitored the temperature of the
wire itself by measuring the increase of its resistance. Passing a
current through the 50\,$\mu$m-wide chip wires, we measured the time
it takes for the wire to reach 120\,C, the results being shown by
the filled circles in Fig.\,\ref{Fig:heating}. In all cases the chip
itself was much colder than the wire. Below 1\,A, the temperature
limit was never reached, but at 1.3\,A, the wire approached 120\,C
in ten seconds. For the 25\,$\mu$m-wide wires, the operation was
continuous below 0.5\,A and was limited to ten seconds at 0.75\,A,
as illustrated by the open squares in Fig.\,\ref{Fig:heating}.

Since a suitable field gradient for the MOT is 0.15\,T/m, the normal
operating current is 5\,mA for a 200\,$\mu$m pyramid and 100\,mA for
a 1 mm pyramid. At these low currents there is negligible heating of
the chip. By contrast, the 1\,mm loop needs to operate at 1\,A if it
is to make a purely magnetic trap for a 100\,$\mu$K cloud of atoms.
This cannot be sustained indefinitely, as shown in
Fig.\,\ref{Fig:heating}, but substantially less than one second
should suffice for most practical applications. Since the field
scales as $I/L$, the situation is even better for purely magnetic
trapping in the smaller loops.

In the course of these measurements, we found that the resistivity
of the gold wire on the chip is 3.9x10$^{-8}\Omega $-m,
approximately 1.6 times higher than that of bulk gold. This is
typical of electro-deposited gold. The main consequence for us is a
slightly higher power dissipation than we had anticipated in our
design using the book value for the resistivity.

\subsection{Optical properties}

Observation of the pyramids under a microscope allowed us to check
that the type-2 and type-3 reflections were indeed eliminated by
removal of the gold coating to create the flower pattern.
Figure\,\ref{Fig:heating}a shows the image using unpolarized light
with the microscope focussed on the apex of an unpatterned pyramid.
In this figure most of the area is bright since the mirrors are
reflecting light over the whole area of the pyramid. By contrast,
Fig.\,\ref{Fig:heating}b shows the same pyramid illuminated with
linearly polarized light and viewed through a crossed polarizer,
which suppresses all but the type-3 contribution, making the corner
regions bright. Figures\,\ref{Fig:heating}c and d show the
corresponding images for a patterned pyramid. In c we see that there
are no longer reflections near the apex of the pyramid, in the
region of the type-2 rays. In d, we see that the removal of the gold
at the edges has completely suppressed the type-3 reflections.

\begin{figure}[t]
\centering
\includegraphics{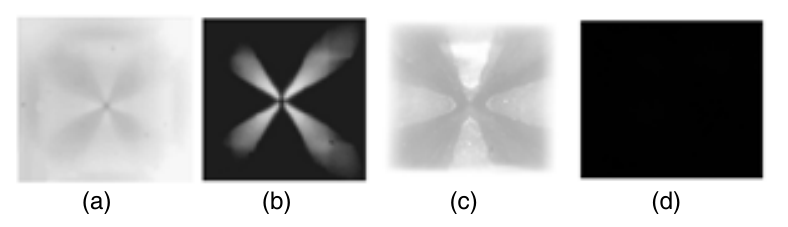}
\caption{Images of the pyramidal mirrors under a microscope,
focussed on the apex. (a \& b) unpatterned pyramids, (c \& d)
patterned pyramids.  (a \& c) No polarizers.  (b \& d) Crossed
polarizer and analyzer.}
\end{figure}

\section{Outlook}\label{Sec:outlook}
\begin{figure}[t]
\centering
\includegraphics{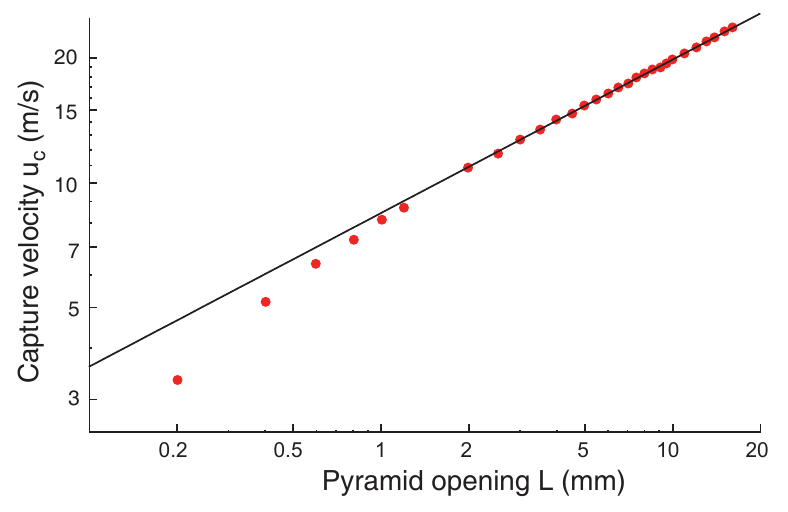}
\label{fig9} \caption{Capture velocity versus the size of the
pyramid opening. Circles: calculation by numerical integration using
the model of \cite{Lindquist:1992}. Line: empirical scaling law $u_c \propto L^{0.37}$. } \label{Fig:uc}
\end{figure}

It is useful to estimate the number of cold atoms that may be
captured in one of these micro-pyramids.  This will depend on the
usual operating variables, such as the laser detuning from the
atomic transition, the positioning of the magnetic field zero, the
balance of the light beams inside the pyramid and the pressure of
the atomic vapour.  In addition, these pyramids are at the extreme
limit of small laser beam size.  Normally the laser beams of a MOT
are one or two centimetres in diameter, whereas these pyramids are
less than a millimetre across.  According to the well-established
model first described by Wieman \textit{et
al.}\,\cite{Lindquist:1992}, the number of atoms $N$ captured in a
MOT is expected to scale as $N\propto L^{2} u_{c}^4$. Here the $L^2$
factor derives from the area of the laser beam, which in our case is
set by the area of the pyramid opening. The quantity $u_c$ is the
capture velocity, i.e. the speed of the fastest atoms captured by
the MOT from the thermal background vapour.  Using the model of
Ref.\,\cite{Lindquist:1992}, we have computed $u_{c}$ numerically,
setting the maximum allowed stopping distance equal to the vertical
height of the pyramid. The circles in Fig.\,\ref{Fig:uc} show the
results on a log-log plot for a variety of pyramid sizes, with the
laser detuning optimised separately to maximise $u_c$ for each size.
For pyramids larger than 1\,mm, we find the empirical scaling law
$u_{c}\propto L^{0.37}$ (the solid line in Fig.\,\ref{Fig:uc}), leading to the result $N\propto L^{3.48}$.
When the pyramids are smaller than that, the atom number drops a
little more rapidly. We have established in a previous
experiment\,\cite{Trupke:2006} that a 16\,mm pyramid made from glass
blocks can capture $10^{8}$ atoms. With the help of this model, we
extrapolate our 16 mm result to predict that a 1\,mm pyramid will
capture some $6\times 10^{3}$ atoms.  The smallest pyramids on this
chip, having $L=0.2\,$mm, are expected by the same argument to
collect approximately 25 atoms.

Although this model is a reasonable estimate, it does neglect some
aspects of the full 3-dimensional geometry.  For example, it assumes
that the atoms of the vapour have the normal thermal distribution
close to the walls of the pyramid and it neglects the polarisation
gradients in the laser field, which lead to additional Sisyphus
cooling\,\cite{Metcalf:99}.  Experiment will have to determine how many atoms are
actually captured. The number of atoms needed depends of course on
the application.  At one extreme, with an array of small clouds,
each containing perhaps $10^{4}$ atoms, the relative displacement of
the clouds could provide a map of local magnetic field variations or
be used to sense inertial forces.  At the other extreme, the
pyramids could serve as single-atom sources for loading integrated
optical cavities, which have recently been demonstrated
\cite{Trupke:2007a}. This would permit the production of single
photons on demand for applications in quantum information
processing.

\section{Summary}
We have fabricated a silicon chip designed to trap cold atoms in an
array of integrated magneto-optical traps. The device contains 48
micro-fabricated hollow micro-pyramids surrounded by electroplated
gold wires. This was packaged into a ceramic pin-grid array, and the
chip's optical and magnetic properties were tested. We found that
the chip wires can easily sustain the currents needed to operate the
MOTs and can even operate for many seconds at the much higher
currents needed to trap atoms magnetically. The use of
electrochemical deposition of photoresist allowed us to pattern a
flower design on the reflective coating, a feature necessary for
achieving the proper optical conditions in the magneto-optical
traps. Numerical estimates show that we should be able to collect
small but useful numbers of atoms in these traps, providing a simple
way to feed arrays of devices on future chips. This represents an
important integration step because atom chips currently have to be
loaded from a single large cloud by a series of somewhat involved
manoeuvres.

\end{document}